\documentclass[twocolumn,aps,prc,superscriptaddress,nofootinbib]{revtex4}

\usepackage{changes}
\usepackage{sidecap}
\usepackage{ulem}
\usepackage{epsfig}
\usepackage{amsmath,amssymb,amsthm}
\usepackage{graphicx}
\usepackage{bm}
\usepackage{color,soul}

\DeclareMathAlphabet{\mathpzc}{OT1}{pzc}{m}{it}

\begin{document}

\newcommand{\old}[1]{\textit{#1}}
\newcommand{\new}[1]{\textbf{#1}}

\renewcommand{\textfraction}{0.00}


\newcommand{\vAi}{{\cal A}_{i_1\cdots i_n}}
\newcommand{\vAim}{{\cal A}_{i_1\cdots i_{n-1}}}
\newcommand{\vAbi}{\bar{\cal A}^{i_1\cdots i_n}}
\newcommand{\vAbim}{\bar{\cal A}^{i_1\cdots i_{n-1}}}
\newcommand{\htS}{\hat{S}}
\newcommand{\htR}{\hat{R}}
\newcommand{\htB}{\hat{B}}
\newcommand{\htD}{\hat{D}}
\newcommand{\htV}{\hat{V}}
\newcommand{\cT}{{\cal T}}
\newcommand{\cM}{{\cal M}}
\newcommand{\cMs}{{\cal M}^*}
\newcommand{\vk}{\vec{\mathbf{k}}}
\newcommand{\bk}{\bm{k}}
\newcommand{\kt}{\bm{k}_\perp}
\newcommand{\kp}{k_\perp}
\newcommand{\km}{k_\mathrm{max}}
\newcommand{\vl}{\vec{\mathbf{l}}}
\newcommand{\bl}{\bm{l}}
\newcommand{\bK}{\bm{K}}
\newcommand{\bb}{\bm{b}}
\newcommand{\qm}{q_\mathrm{max}}
\newcommand{\vp}{\vec{\mathbf{p}}}
\newcommand{\bp}{\bm{p}}
\newcommand{\vq}{\vec{\mathbf{q}}}
\newcommand{\bq}{\bm{q}}
\newcommand{\qt}{\bm{q}_\perp}
\newcommand{\qp}{q_\perp}
\newcommand{\bQ}{\bm{Q}}
\newcommand{\vx}{\vec{\mathbf{x}}}
\newcommand{\bx}{\bm{x}}
\newcommand{\tr}{{{\rm Tr\,}}}
\newcommand{\bc}{\textcolor{blue}}

\newcommand{\beq}{\begin{equation}}
\newcommand{\eeq}[1]{\label{#1} \end{equation}}
\newcommand{\ee}{\end{equation}}
\newcommand{\bea}{\begin{eqnarray}}
\newcommand{\eea}{\end{eqnarray}}
\newcommand{\beqar}{\begin{eqnarray}}
\newcommand{\eeqar}[1]{\label{#1}\end{eqnarray}}

\newcommand{\half}{{\textstyle\frac{1}{2}}}
\newcommand{\ben}{\begin{enumerate}}
\newcommand{\een}{\end{enumerate}}
\newcommand{\bit}{\begin{itemize}}
\newcommand{\eit}{\end{itemize}}
\newcommand{\ec}{\end{center}}
\newcommand{\bra}[1]{\langle {#1}|}
\newcommand{\ket}[1]{|{#1}\rangle}
\newcommand{\norm}[2]{\langle{#1}|{#2}\rangle}
\newcommand{\brac}[3]{\langle{#1}|{#2}|{#3}\rangle}
\newcommand{\hilb}{{\cal H}}
\newcommand{\pleft}{\stackrel{\leftarrow}{\partial}}
\newcommand{\pright}{\stackrel{\rightarrow}{\partial}}
\newcommand{\dif}{\mathrm{d}}
\newcommand{\pT}{p_\perp}
\newcommand{\sNN}{\sqrt{s_{\mathrm{NN}}}}
\newcommand{\RAA}{R_{\mathrm{AA}}}
\newcommand{\lton}{\mathrel{\lower.9ex \hbox{$\stackrel{\displaystyle
<}{\sim}$}}}


\title{Early evolution constrained by high-$\pT$ QGP tomography}

\author{Stefan Stojku}
\affiliation{Institute of Physics Belgrade, University of Belgrade, Serbia}

\author{Jussi Auvinen}
\affiliation{Institute of Physics Belgrade, University of Belgrade, Serbia}

\author{Marko Djordjevic}
\affiliation{Faculty of Biology, University of Belgrade, Serbia}

\author{Pasi Huovinen\footnote{E-mail: pasi@ipb.ac.rs}}
\affiliation{Institute of Physics Belgrade, University of Belgrade, Serbia}

\author{Magdalena Djordjevic\footnote{E-mail: magda@ipb.ac.rs}}
\affiliation{Institute of Physics Belgrade, University of Belgrade, Serbia}

\begin{abstract}
  We show that high-$\pT$ $\RAA$ and $v_2$ are sensitive to the early
  expansion dynamics, and that the high-$\pT$ observables prefer
  delayed onset of energy loss and transverse expansion. To calculate
  high-$\pT$ $\RAA$ and $v_2$, we employ our newly developed DREENA-A
  framework, which combines state-of-the-art dynamical energy loss
  model with 3+1-dimensional hydrodynamical simulations. The model
  applies to both light and heavy flavor, and we predict a larger
  sensitivity of heavy flavor observables to the onset of transverse
  expansion. This presents the first time when bulk QGP behavior has
  been constrained by high-$p_\perp$ observables and related theory,
  i.e., by so-called QGP tomography.
\end{abstract}


\maketitle

Quark-Gluon-Plasma (QGP)~\cite{Collins,Baym} is an extreme form of
matter that consists of interacting quarks, antiquarks and
gluons. This state of matter is formed in ultrarelativistic heavy-ion
collisions at the Relativistic Heavy-Ion Collider (RHIC) and the Large
Hadron Collider (LHC).  When analyzing the heavy-ion collision data,
the particles formed in these collisions are traditionally separated
into high-$\pT$ (rare hard probes) and low-$\pT$ particles (bulk,
consisting of $99.9\%$ of particles formed in these collisions).

The QGP properties are traditionally explored by low-$\pT$
observables~\cite{Teaney,Shen:2020gef,QGP1,QGP2}, while rare
high-$\pT$ probes are, almost exclusively, used to understand the
interactions of high-$\pT$ partons with the surrounding QGP medium.
High-$\pT$ physics had a decisive role in the QGP
discovery~\cite{QGP3}, but it has been rarely used to understand bulk QGP
properties. On the other hand, some important bulk QGP properties are
difficult to constrain by low-$p_\perp$ observables and
corresponding theory/simulations~\cite{Nagle,Koop,Auvinen,Niemi}. We
are therefore advocating QGP tomography, where bulk QGP
parameters are {\it jointly} constrained by low- and high-$\pT$
physics.

During the last few years, our understanding of the very early
evolution of QGP has evolved a lot. In particular the discovery of the
attractor solutions of the evolution of non-equilibrated
systems~\cite{Heller:2015dha,Akamatsu:2020lej,Shen:2020mgh}, and
models based on effective kinetic
theory~\cite{Kurkela:2018wud,Kurkela:2018vqr} have been significant
milestones. However, the exact dynamics of early evolution and
hydrodynamization of the medium---i.e. the approach to the state where
the system can be described using fluid dynamics---are not settled
yet. Furthermore, to our knowledge, there are no reliable methods to
calculate jet energy loss in a medium out of equilibrium. Instead of
microscopic calculation of the early-time dynamics, we take a
complementary approach in this paper. We calculate the high-$\pT$
$\RAA$ and $v_2$ in a few straightforward scenarios, and show how the
comparison to high-$\pT$ data constrains the early evolution.

In the attractor solutions, the final evolution is fluid dynamical
even if the initial state is quite far from equilibrium. This allows
us to entertain the notion that even if the early state is not in
local equilibrium, we could use fluid dynamics to describe its
evolution from very early times~\cite{Chattopadhyay:2019jqj}, say from
$\tau_0 = 0.2$ fm, where $\tau_0$ is the initial time of fluid
dynamical evolution.  Correspondingly, we may argue that the
temperature entering fluid dynamical evolution controls also jet
energy loss, and we may start the jet energy loss at the same time,
$\tau_q = 0.2$ fm. On the other hand, we had studied the
pre-equilibrium energy loss in various scenarios~\cite{Zigic:2019sth},
and seen that even if the data could not properly distinguish these
scenarios, Bjorken-type temperature evolution at very early times
tended to push $\RAA$ too low. This may suggest that applying the
equilibrium jet-medium interactions to the pre-equilibrium stage (even
if close enough to fluid dynamical) overestimates the energy loss. Due
to this, we here, for simplicity, assume an opposite limit, where we
start the energy loss later than the fluid dynamical evolution:
$\tau_q = 1.0$ fm and $\tau_0 = 0.2$ fm~\footnote[1]{Similar scenario was
  suggested and studied in Ref.~\cite{Andres}.}.

Frequently used toy model to study the effects of early
non-equilibrium evolution is the free streaming
approach~\cite{Broniowski:2008qk,Liu:2015nwa}, where (fictional)
particles are allowed to stream freely until the initial time of fluid
dynamical evolution $\tau_0$. As our third scenario, we allow free
streaming until $\tau_0 = 1.0$ fm. Consistently with the assumed
absence of interactions in the bulk medium, we assume no jet-medium
interactions during the out-of-equilibrium stage, so that
$\tau_0=\tau_q = 1.0$~fm.  For comparison's sake, we also explore the
``old-fashioned'' scenario where ``nothing'' happens before the fluid
dynamical initial time $\tau_0 = \tau_q = 1.0$ fm, i.e.\ we start the
fluid-dynamical evolution at $\tau = 1.0$ fm with zero transverse flow
velocity.

\begin{figure}[h]
 \includegraphics[width=8cm]{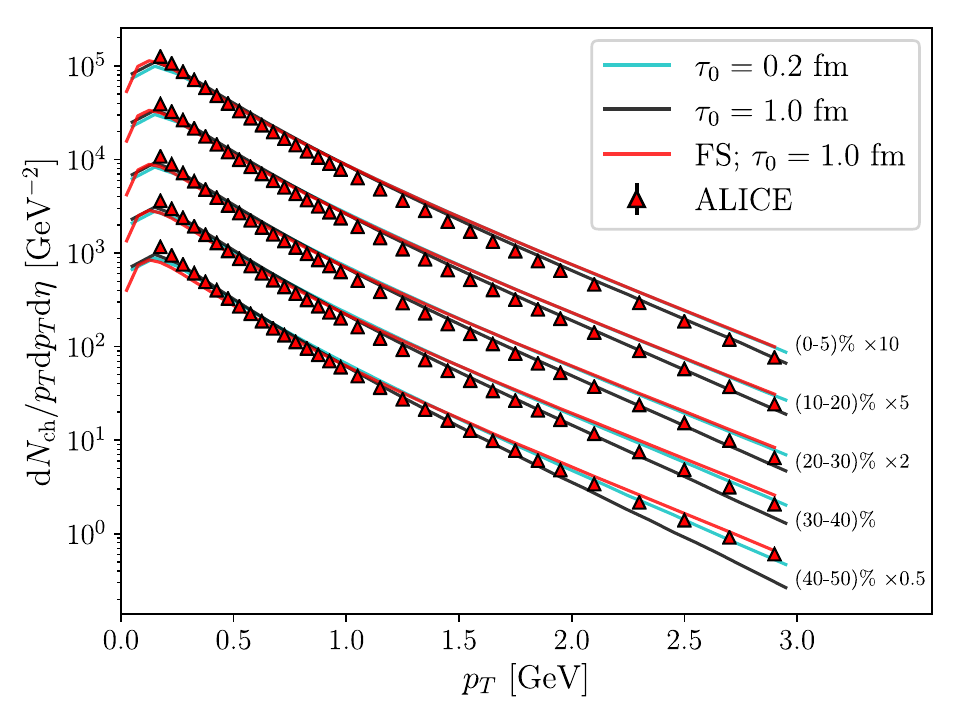}
 \vspace*{-0.5cm}
    \caption{Transverse momentum spectrum of charged particles in five
      centrality classes in Pb+Pb collisions at $\sqrt{s_{NN}}=5.02$
      TeV, with two initial times $\tau_0 = 0.2$ and $\tau_0 = 1.0$
      fm, and free streaming initialization (FS). ALICE data
      from~\cite{ALICE_CH_RAA}.}
\label{fig1}
\end{figure}

\begin{figure*}
 \includegraphics[width=\textwidth]{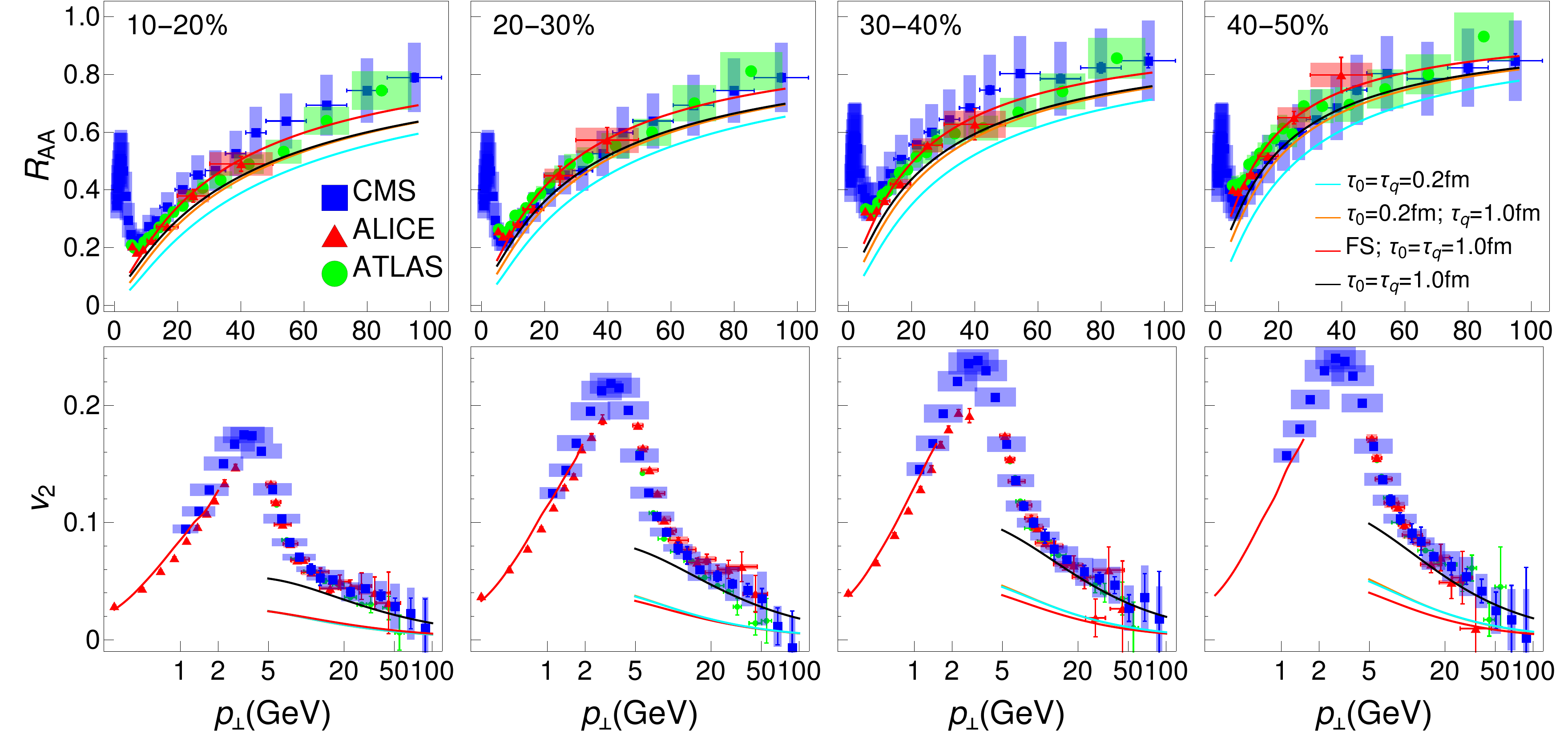}
 \caption{Charged hadron DREENA-A $R_{AA}$ (upper panels) and $v_2$
   (lower panels) predictions, generated for different $\tau_0$,
   $\tau_q$, and initialization (see the legend, FS stands for free
   streaming), are compared with ALICE~\cite{ALICE_CH_RAA,ALICE_CH_v2},
   CMS~\cite{CMS_CH_RAA,CMS_CH_v2} and ATLAS~\cite{ATLAS_CH_RAA,ATLAS_CH_v2}
   data. Four columns, from left to right, correspond to 10--20\%, 20--30\%,
   30--40\% and 40--50\% centralities at $\sqrt{s_{NN}}=5.02$ Pb+Pb collisions
   at the LHC. At low $\pT$ ($\pT < 2$ GeV) $v_2$ is 4-cumulant $v_2\{4\}$,
   whereas at high $\pT$ ($\pT > 5$ GeV) we evaluate $v_2$ as
   $v_2 = (1/2)\, (\RAA^{\mathrm{in}} - \RAA^{\mathrm{out}})
               /(\RAA^{\mathrm{in}} + \RAA^{\mathrm{out}})$.}
\label{fig2}
\end{figure*}

When calculating how the high-$\pT$ observables depend on our
different scenarios we have to ensure that the QGP medium evolution is
compatible with the observed distributions of low-$\pT$ particles. We
describe the medium evolution using the 3+1-dimensional viscous
hydrodynamical model~\cite{Molnar:2014zha}. For simplicity,
we choose a constant shear viscosity to entropy density ratio
$\eta/s=0.12$ for the cases without pre-hydro transverse flow, and
$\eta/s=0.16$ for the free-streaming initialization. In all the cases
the initial energy density profile in transverse plane is given by the
binary collision density $n_{BC}$ from the optical Glauber model:
\beq
e(\tau_0, x, y, b) = C_e(\tau_0)\left(n_{BC} + c_1 n_{BC}^2 + c_2 n_{BC}^3\right).
\ee
The parameters $C_e$, $c_1$ and $c_2$ are tuned separately for each
scenario, to approximately describe the observed charged particle
multiplicities and $v_2\{4\}$ in Pb+Pb collisions at
$\sqrt{s_{NN}}=5.02$ TeV. For the longitudinal profile, we keep the
parametrization used for $\sqrt{s_{NN}}=2.76$ Pb+Pb collisions~\cite{Molnar:2014zha}. The equation of state is
$s95p$-PCE-v1~\cite{Huovinen:2009yb}. We use freeze-out
temperatures $T_{\mathrm{chem}} = 150$ MeV and $T_{\mathrm{dec}} = 100$ MeV
for cases without pre-hydro flow, but with free streaming we use
$T_{\mathrm{chem}} = 175$ MeV~\cite{Niemi:2015qia} to mimic bulk
viscosity around $T_c$ required to fit the $p_T$ distributions, and
$T_{\mathrm{dec}} = 140$ MeV.

In the free-streaming initialization massless particles stream freely
from $\tau = 0.2$ fm to $\tau_0 = 1.0$ fm, where the energy-momentum
tensor based on the distributions of these particles is evaluated. The
energy momentum tensor is decomposed to densities, flow velocity and
dissipative currents, which are used as the initial state of the
subsequent fluid-dynamical evolution.  The switch from massless
non-interacting particles to strongly interacting constituents of QGP
causes large positive bulk pressure at $\tau_0$. In our calculations
bulk viscosity coefficient is always zero, and the initial bulk
pressure will approach zero according to Israel-Stewart equations.

The transverse momentum distributions of charged particles are shown
in Fig.~\ref{fig1}, and $\pT$-differential elliptic flow parameter
$v_2\{4\}(\pT)$ in the low momentum part ($\pT < 2$ GeV) of the lower
panels of Fig.~\ref{fig2}. As seen, the overall agreement with the
data is acceptable.

To be able to use the high-$\pT$ sector to study the bulk behavior
we need a framework that incorporates both state-of-the-art
energy loss and bulk medium simulations. With this goal, we recently
developed a fully optimized modular framework
DREENA-A~\cite{DREENAA}, which can incorporate any, arbitrary, temperature
profile within the dynamical energy loss formalism (outlined
below). Consequently, ``DREENA'' stands for Dynamical Radiative and
Elastic ENergy loss Approach, while ``A'' stands for Adaptive. The
framework does not have fitting parameters within the energy loss model,
allowing to fully exploit different temperature profiles (as the only
input in the DREENA-A framework), systematically compare the data and
predictions obtained by the same formalism and parameter set, and
consequently constrain the bulk QGP properties from jointly studying
low and high-$\pT$ theory and data.

The initial quark spectrum is computed at next to leading
order~\cite{Vitev0912} for light and heavy partons. To generate
charged hadrons, we use DSS~\cite{DSS} fragmentation functions. For D
and B mesons, we use BCFY~\cite{BCFY} and KLP~\cite{KLP} fragmentation
functions, respectively. In the presence of QCD medium, the
vacuum fragmentation functions should be modified along with parton
energy loss as described by the multi-scale
models~\cite{JETSCAPE:2021ehl,Ke:2020clc}. However, for
high-$\pT>10$~GeV, which is the momentum region covered in our
study~\footnote[2]{As the assumptions in the dynamical energy loss break
  down below 10~GeV, we consider our predictions to be reliable in the
  region $\pT>10$~GeV.}, such modification is small, justifying the
use of vacuum fragmentation~\cite{JETSCAPE:2021ehl}.

The dynamical energy loss formalism~\cite{MD_Dyn,MD_Coll} has several
unique features: {\it i)} QCD medium of {\it finite} size and
temperature consisting of dynamical (i.e.\ moving) partons; this in
distinction to medium models with widely used static approximation
and/or vacuum like propagators~\cite{BDMPS,ASW,GLV,HT}.
{\it ii)} Calculations based on generalized Hard-Thermal-Loop
approach~\cite{Kapusta}, with naturally regulated infrared
divergences~\cite{MD_Dyn,MD_Coll,DG_TM}.
{\it iii)} Calculations of both radiative~\cite{MD_Dyn} and
collisional~\cite{MD_Coll} energy loss in the same theoretical
framework.
{\it iv)} Generalization towards running coupling~\cite{MD_PLB},
finite magnetic mass~\cite{MD_MagnMass}. We also recently advanced the 
formalism towards relaxing the widely used soft-gluon approximation~\cite{sga}.
All of these features are necessary for accurate
predictions~\cite{Blagojevic_JPG}, but utilizing evolving temperature
profiles is highly non-trivial within this complex energy loss framework.

We use the same parameter set to generate high-$p_\perp$ predictions
as in our earlier studies within DREENA-C~\cite{DREENAc} and
DREENA-B~\cite{DREENAb} frameworks. In particular, we use
$\Lambda_{QCD}=0.2$~GeV and effective light quark flavors $n_f{\,=\,}3$.
For light quark mass, we assume to be dominated by the thermal mass
$M{\,=\,}\mu_E/\sqrt{6}$, and for the gluon mass, we take
$m_g=\mu_E/\sqrt{2}$~\cite{DG_TM}. The temperature-dependent Debye mass
$\mu_E$ is obtained by applying procedure from~\cite{Peshier}, which
leads to results compatible with the lattice QCD~\cite{LatticeMass}.
The charm (bottom) mass is $M{\,=\,}1.2$\,GeV  ($M{\,=\,}4.75$\,GeV).
Magnetic to electric mass ratio is $0.4 < \mu_M/\mu_E < 0.6$~\cite{Maezawa,Nakamura,Hart,Bak}, but for simplicity $\mu_M/\mu_E = 0.5$, leading to the uncertainty of up to $10\%$ for both $R_{AA}$ and $v_2$ results.

The resulting DREENA-A predictions
for charged hadron $R_{AA}$ and $v_2$ in four different
centrality classes, and four scenarios of early evolution, are shown
in Fig.~\ref{fig2}, and compared with experimental data. As one can
expect, the later the energy loss begins, the higher the $\RAA$, and
evaluating the energy loss as in thermalized medium already at $\tau_q = 0.2$ fm
is slightly disfavored. Furthermore, early free-streaming evolution
leads to larger $\RAA$ than fluid-dynamical evolution. On the other
hand, the behavior of $v_2$ is different. First, if the
early expansion is fluid dynamical, we see that delaying the onset of
energy loss hardly changes $v_2$ at all. Second, early free-streaming
evolution does not lead to better reproduction of the data, but,
in peripheral collisions, the fit is even worse. The only case when
our $v_2$ predictions approach the data, is when {\it both} the jet
energy loss {\it and} the transverse expansion are delayed to $\tau = 1$ fm.

\begin{figure}
 \includegraphics[width=8.5cm]{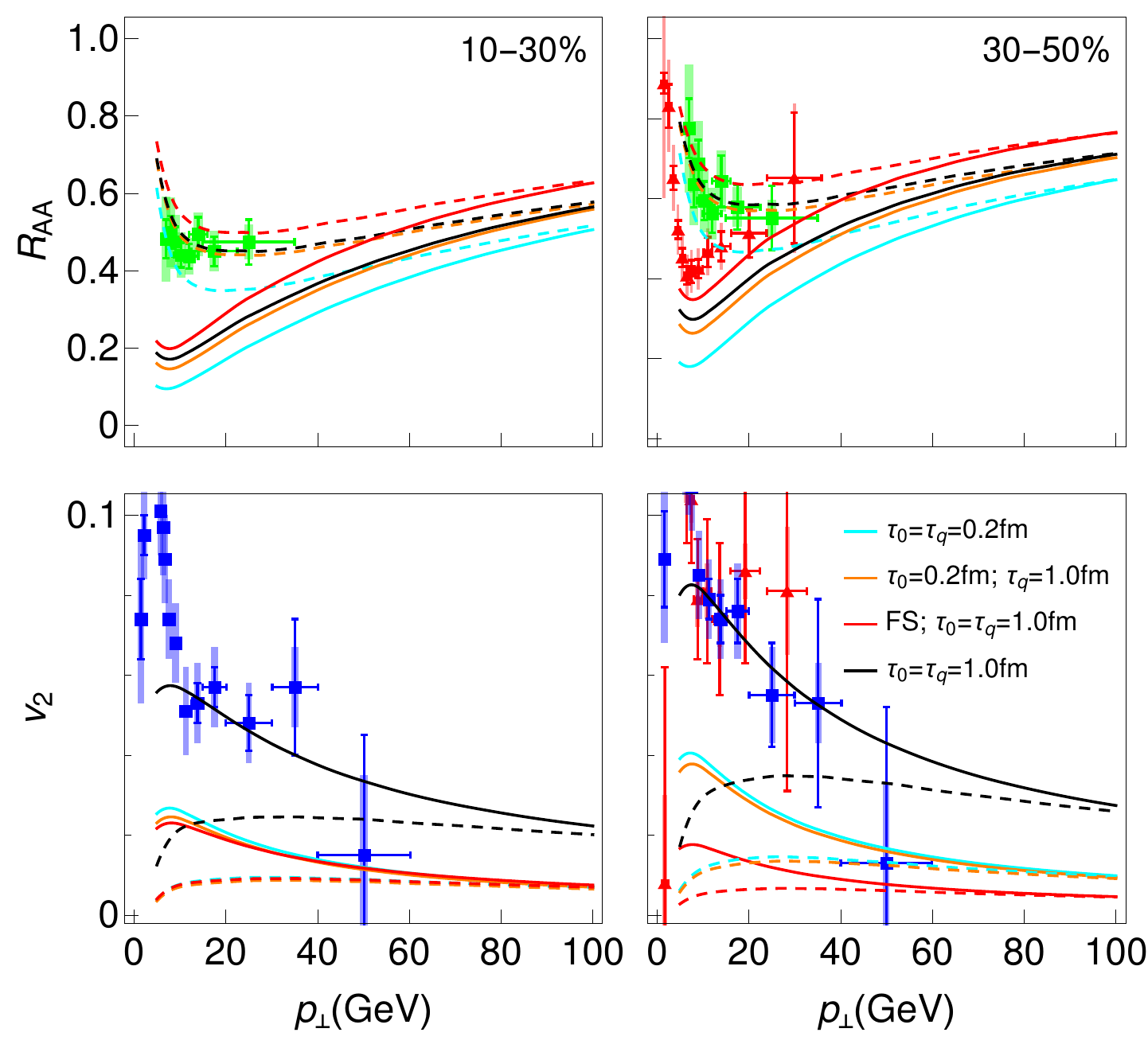}
 \caption[width=8.5cm]{Predicted D (full curves) and B meson (dashed
   curves) $R_{AA}$ (upper panels) and $v_2$ (lower panels) in Pb+Pb
   collisions at $\sqrt{s_{NN}}=5.02$ TeV. The predictions for D
   mesons are compared with ALICE~\cite{ALICE_D_RAA,ALICE_D_v2} (red
   triangles) and CMS~\cite{CMS_D_v2} (blue squares) D meson data,
   while predictions for B mesons are compared with
   CMS~\cite{CMS_JPsi} (green circles) non-prompt $J/\Psi$ data.}
\label{fig3}
\end{figure}
\begin{figure}
 \includegraphics[width=8.5cm]{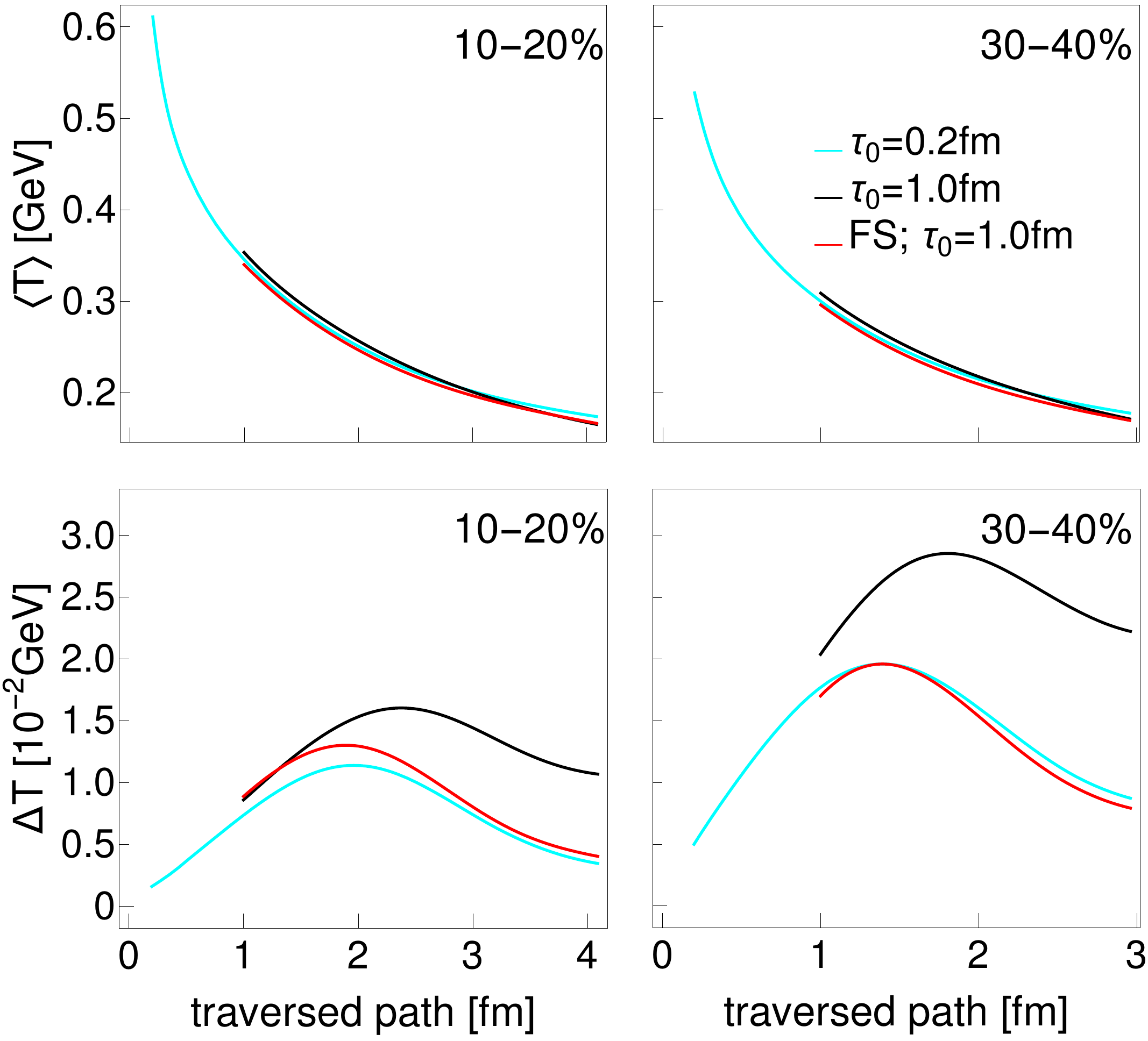}
 \caption{Average temperature along the jet path traversing the system
   (upper panel) and the difference of average temperatures in
   out-of-plane and in-plane directions (lower panel) for $\tau_0 = 0.2$
   and 1.0 fm and free-streaming initialization at 10-20\% and 30-40\%
   centrality classes. The average is over all sampled jet paths, and
   the path ends at $T_C \approx 160$~MeV~\cite{Tc}.}
 \label{fig5}
\end{figure}

As shown in Fig.~\ref{fig3}, heavy quarks are even more sensitive to
the early evolution. For bottom probes, the data are largely not
available, making these true predictions. For charm probes, the
available experimental data are much more sparse (and with larger
error bars) than the charged hadron data. However, where available,
comparison of our predictions with the data suggests the same
preference towards delayed energy loss and transverse expansion as
charged hadrons. These results are important, as consistency between
light and heavy flavor is crucial (though highly non-trivial, as
e.g. implied by the well known heavy flavor puzzle~\cite{HFPuzzle}) for
studying the QGP properties.

To investigate the origin of the sensitivity of $\RAA$ and $v_2$ to
the early evolution, we evaluate the temperature along the paths of
jets traveling in-plane ($\phi = 0$) and out-of-plane ($\phi = \pi/2$)
directions, and average over all sampled jet paths. In Fig.~\ref{fig5}
we show the time evolution of the average of temperatures in in- and
out-of-plane directions, and their difference in 10--20\% and 30-40\%
central collisions for $\tau_0 = 0.2$ and 1.0 fm, and the free
streaming initialization. The behavior of $\RAA$ is now easy to
understand in terms of average temperature: Larger $\tau_q$,
i.e.\ delay in the onset of energy loss, cuts away the large
temperature part of the profile decreasing the average temperature,
and thus increasing the 
$R_{AA}$~\cite{DREENAc,DREENAb}.
Similarly, for late start of transverse expansion, i.e.\ $\tau_0 = 1.0$ fm,
the temperature is first slightly larger and later lower than for
$\tau_0 = 0.2$ fm, and thus the $\RAA$ in $\tau_0 = \tau_q = 1.0$ fm
and $\tau_0 = 0.2$ with $\tau_q = 1.0$ cases is almost identical. On the
other hand, due to the rapid expansion of the edges of the system,
free streaming initialization leads to lower average temperature than
any other scenario, and thus to the largest $\RAA$.

High-$\pT$ $v_2$, on the other hand, is proportional to the difference
in temperature along in-plane and out-of-plane directions, and to
lesser extent to the average temperature. Delaying the onset of
transverse expansion to $\tau_0 = 1.0$ fm leads to larger difference
than either early fluid-dynamical or free streaming expansion, and
thus $v_2$ is largest in that case. As well, delaying the onset of
energy loss by increasing $\tau_q$ hardly changes $v_2$, since at
early times the temperature seen by jets in in- and out-of-plane
directions is almost identical, and no $v_2$ is built up at that time.
Early free streaming and early fluid-dynamical expansion lead to
similar differences in temperatures. The slightly larger difference
in the 10-20\% centrality class is counteracted by slightly lower
temperature, and thus final $v_2$ is practically identical in both
cases. In the more peripheral 30-40\% class the differences in
temperature are almost identical, but the lower average temperature
leads to lower $v_2$ for free streaming.

The delay in transverse expansion affects the average temperature
along the jet in two ways. First, smaller $\tau_0$ means larger
initial gradients, faster build-up of flow, and faster dilution of the
initial spatial anisotropy. Similarly, free-streaming leads
to even faster build-up of flow and dilution of spatial anisotropies
than early fluid-dynamical expansion. Second, since the initial jet
production is azimuthally symmetric, and jets travel along eikonal
trajectories, at early times both in- and out-of-plane jets probe the
temperature of the medium almost the same way. Only with course of
time will the spatial distribution of in- and out-of-plane jets
differ, and the average temperature along their paths begins to
reflect the anisotropies of the fluid temperature.  This qualitative
understanding indicates that the obtained
conclusions are largely model independent.

The idea of using high-$\pT$ theory and data to explore
QGP is not new, see e.g.~Refs.~\cite{Vitev:2002pf,Renk:2010qx,
  Betz:2014cza,Andres,Shi:2018izg,Shi:2018lsf,JET:2013cls,Kumar:2019uvu,
  LBT1,LBT2,Xu:2014ica}. While some of these approaches can achieve a reasonable agreement with the
data (see e.g.~\cite{Shi:2018izg,Zhao:2021vmu,Werner:2012xh}), this agreement relies on adjusting fitting parameter(s) in the energy loss model, which prevents them from constraining the bulk medium properties. These models thus largely concentrate on investigating the nature of parton interactions (e.g., a new phenomenon of magnetic monopoles is systematically introduced in~\cite{Shi:2018izg}) rather than exploring which dynamical evolution better explains the data. In contrast, the goal of our approach is to constrain the bulk QGP behavior. The major advantage of our framework is that it does not use fitting parameters in the energy loss model, enabling us to explore the effects of different bulk medium evolutions. We can even use $R_{AA}$ to make conclusions about the bulk properties of the system, where our $R_{AA}$ results imply that the energy loss during the very early evolution is weaker than energy loss in a fully thermal system.

Furthermore, our study shows that not only is early energy loss
suppressed~\cite{Renk:2010qx,Andres}, but the early build-up of
transverse expansion must be delayed as well. It is not sufficient to
delay cooling as suggested in Ref.~\cite{Renk:2010qx}, but the initial
anisotropy must be diluted at much slower rate than given by either
free streaming or by fluid dynamics. We do not expect current more
sophisticated approaches to pre-equilibrium dynamics, like
K\o{}MP\o{}ST based on effective kinetic
theory~\cite{Kurkela:2018wud,Kurkela:2018vqr}, to resolve this
issue. As seen in Ref.~\cite{NunesdaSilva:2020bfs}, except in most
peripheral collisions, both K\o{}MP\o{}STing and free streaming lead
to very similar final distributions. Thus we may expect that at the
time of switching to fluid dynamics, they both have lead to very
similar flow and temperature profiles (and thus anisotropies).

Alternatively, the initial spatial anisotropies could be way larger
than considered here. It is known that both IP-Glasma and EKRT
approaches lead to larger eccentricities than Glauber, but we have
tested that they both lead to too low high-$\pT$ $v_2$, if the fluid
dynamical evolution begins as usually assumed in calculations
utilizing IP-Glasma or EKRT initializations.  Event-by-event
fluctuations may enhance spatial anisotropies as well, and by
generating shorter scale structures, they may enhance the sensitivity
of high-$\pT$ $v_2$ to spatial anisotropies. However, for these
additional structures to enhance the high-$\pT$ $v_2$, they should be
correlated with the event plane, which is not necessarily the
case. While we have postponed a study of event-by-event fluctuations
to a further work, our preliminary results do not indicate substantial
influence on high-$p_\perp$ predictions.

In summary, we presented (to our knowledge) the first example of using
high-$\pT$ theory and data to provide constraints to bulk QGP
evolution. Specifically, we inferred that experimental data suggest
that at early times both the energy loss and transverse expansion of
the system should be significantly weaker than in conventional models.
We emphasize that the assumption that no energy loss nor transverse
expansion takes place before $\tau_0 = 1.0$ fm is unrealistic.
We are not advocating such a scenario, but note that the only way
available to us to test our hypothesis that the early energy loss and
expansion should be suppressed, was to take the limit of no
energy loss nor transverse expansion at all. Doing this significantly
improves the agreement with the data, thus supporting our hypothesis.
While our finding of delayed onset of energy loss and transverse
expansion has yet to be physically understood, there have been several
anomalies in the history of heavy-ion physics, and our result is one
more of them.

Furthermore, heavy flavor observables show large sensitivity to the
details of early evolution, so our conclusion will be further tested
by the upcoming high luminosity measurements. Our results
demonstrate inherent interconnections between low- and high-$\pT$
physics, strongly supporting the utility of our QGP tomography
approach, where bulk QGP properties are {\it jointly} constrained by
low- and high-$\pT$ data.

{\em Acknowledgments:}
This work is supported by the European
Research Council, grant ERC-2016-COG: 725741, and by the
Ministry of Science and Technological Development of the Republic of
Serbia, under project numbers ON171004 and ON173052.


\begin{references}

\bibitem{Collins}
  J.~C.~Collins and M.~J.~Perry, Phys.~Rev.~Lett. {\bf 34}, 1353 (1975).

\bibitem{Baym} G.~Baym and S.~A.~Chin, Phys.~Lett.~B {\bf 62}, 241 (1976).

\bibitem{Teaney} D.~A.~Teaney, "Viscous hydrodynamics and the quark
  gluon plasma." Quark-Gluon Plasma {\bf 4}, 207 (2011).

\bibitem{Shen:2020gef}
  C.~Shen,
  Nucl. Phys. A \textbf{1005}, 121788 (2021).

\bibitem{QGP1} B.~Jacak and P.~Steinberg, Phys.~Today {\bf 63}, 39 (2010).

\bibitem{QGP2} C.~V.~Johnson and P.~Steinberg, Phys.~Today {\bf 63}, 29 (2010).

\bibitem{QGP3} M.~Gyulassy and L.~McLerran, Nucl.\ Phys.\ A {\bf 750}, 30 (2005).

\bibitem{Nagle}
 J.~L.~Nagle, I.~G.~Bearden, and W.~A.~Zajc,
 New J.~Phys. {\bf 13}, 075004 (2011).

\bibitem{Koop}
  J.~Koop, A.~Adare, D.~McGlinchey, and J.~Nagle, Phys.~Rev. C {\bf 92}, 054903 (2015).

\bibitem{Auvinen}
 J.~Auvinen, J.~E.~Bernhard, S.~A.~Bass and I.~Karpenko,
 Phys.\ Rev.\ C \textbf{97}, 044905 (2018).

\bibitem{Niemi}
 J.~Auvinen, K.~J.~Eskola, P.~Huovinen, H.~Niemi, R.~Paatelainen and P.~Petreczky,
 Phys. Rev. C \textbf{102}, 044911 (2020).

\bibitem{Heller:2015dha}
 M.~P.~Heller and M.~Spalinski,
 Phys. Rev. Lett. \textbf{115}, 072501 (2015).

\bibitem{Akamatsu:2020lej}
 Y.~Akamatsu,
 Nucl. Phys. A \textbf{1005}, 122000 (2021).

\bibitem{Shen:2020mgh}
 C.~Shen and L.~Yan,
 Nucl. Sci. Tech. \textbf{31}, no.12, 122.

\bibitem{Kurkela:2018wud}
 A.~Kurkela, A.~Mazeliauskas, J.~F.~Paquet, S.~Schlichting and D.~Teaney,
 Phys. Rev. Lett. \textbf{122}, 122302 (2019).

\bibitem{Kurkela:2018vqr}
 A.~Kurkela, A.~Mazeliauskas, J.~F.~Paquet, S.~Schlichting and D.~Teaney,
 Phys. Rev. C \textbf{99}, 034910 (2019).

\bibitem{Chattopadhyay:2019jqj}
 C.~Chattopadhyay and U.~W.~Heinz,
 Phys. Lett. B \textbf{801}, 135158 (2020).

\bibitem{Zigic:2019sth}
 D.~Zigic, B.~Ilic, M.~Djordjevic and M.~Djordjevic,
 Phys.\ Rev.\ C \textbf{101}, 064909 (2020).

\bibitem{Broniowski:2008qk}
 W.~Broniowski, W.~Florkowski, M.~Chojnacki and A.~Kisiel,
 Phys. Rev. C \textbf{80}, 034902 (2009).

\bibitem{Liu:2015nwa}
  J.~Liu, C.~Shen and U.~Heinz,
  Phys.~Rev.~C \textbf{91}, 064906 (2015)
  [erratum: Phys. Rev. C \textbf{92}, 049904 (2015)].

\bibitem{Molnar:2014zha}
  E.~Molnar, H.~Holopainen, P.~Huovinen and H.~Niemi,
  Phys.~Rev.~C \textbf{90}, 044904 (2014).

\bibitem{Huovinen:2009yb}
  P.~Huovinen and P.~Petreczky,
  Nucl.\ Phys.\ A \textbf{837}, 26-53 (2010).

\bibitem{Niemi:2015qia}
 H.~Niemi, K.~J.~Eskola and R.~Paatelainen,
 Phys. Rev. C \textbf{93}, 024907 (2016).

\bibitem{ALICE_CH_RAA} S.~Acharya {\it et al.} [ALICE], JHEP {\bf 1811}, 013 (2018).

\bibitem{DREENAA} D.~Zigic, I.~Salom, J.~Auvinen, P.~Huovinen and M.~Djordjevic,
[arXiv:2110.01544 [nucl-th]].

\bibitem{Vitev0912} Z. B. Kang, I. Vitev and H. Xing, Phys. Lett. B {\bf 718}, 482 (2012), R. Sharma, I. Vitev and B.W. Zhang, Phys. Rev. C {\bf 80}, 054902  (2009).

\bibitem{DSS} D. de Florian, R. Sassot and M. Stratmann, Phys. Rev. D {\bf 75}, 114010 (2007).

\bibitem{BCFY} M. Cacciari, P. Nason, JHEP {\bf 0309}, 006 (2003), E. Braaten, K.-M. Cheung, S. Fleming and T. C. Yuan, Phys. Rev. D {\bf 51}, 4819 (1995).

\bibitem{KLP} V. G. Kartvelishvili, A.K. Likhoded, V.A. Petrov, Phys. Lett. B {\bf 78}, 615 (1978).

\bibitem{JETSCAPE:2021ehl}
S.~Cao \textit{et al.} [JETSCAPE],
Phys. Rev. C \textbf{104} (2021) no.2, 024905.

\bibitem{Ke:2020clc}
W.~Ke and X.~N.~Wang,
JHEP \textbf{05} (2021), 041.

\bibitem{MD_Dyn}  M.~Djordjevic, Phys.\ Rev.\ C {\bf 80}, 064909 (2009);
  M.~Djordjevic and U.~Heinz, Phys.\ Rev.\ Lett.\ {\bf 101}, 022302 (2008).

\bibitem{MD_Coll}  M.~Djordjevic, Phys.\ Rev.\ C {\bf 74}, 064907 (2006).

\bibitem{BDMPS} R.~Baier, Y.~Dokshitzer, A.~Mueller, S.~Peigne, and
  D.~Schiff, Nucl.\ Phys.\ B \textbf{484}, 265 (1997).

\bibitem{ASW} N.~Armesto, C.~A.~Salgado, and U.~A.~Wiedemann,
  Phys.\ Rev.\ D \textbf{69}, 114003 (2004).

\bibitem{GLV} M.~Gyulassy, P.~L\'evai, and I.~Vitev,
  Nucl.\ Phys.\ B \textbf{594}, 371 (2001).

\bibitem{HT}
  X.~N.~Wang and X.~f.~Guo,
  Nucl.\ Phys.\ A \textbf{696}, 788-832 (2001).

\bibitem{Kapusta}
  J.~I.~Kapusta, {\it Finite-Temperature Field Theory}
  (Cambridge University Press, 1989).

\bibitem{DG_TM} M.~Djordjevic and M.~Gyulassy, Phys.\ Rev.\ C {\bf 68}, 034914 (2003).

\bibitem{MD_PLB}
  M.~Djordjevic and M. Djordjevic, Phys.\ Lett.\ B {\bf 734}, 286 (2014).

\bibitem{MD_MagnMass}
  M.~Djordjevic and M.~Djordjevic, Phys.\ Lett.\ B {\bf 709}, 229 (2012).

\bibitem{sga}
  B.~Blagojevic, M.~Djordjevic and M.~Djordjevic, Phys.\ Rev.\ C {\bf 99}, 024901 (2019).

\bibitem{Blagojevic_JPG}
  B.~Blagojevic and M.~Djordjevic, J.\ Phys.\ G {\bf 42}, 075105 (2015).

\bibitem{DREENAc}  D.~Zigic, I.~Salom, J.~Auvinen, M.~Djordjevic and M.~Djordjevic,
  J.\ Phys.\ G {\bf 46}, 085101 (2019).

\bibitem{DREENAb} D.~Zigic, I.~Salom, J.~Auvinen, M.~Djordjevic and M.~Djordjevic,
  Phys.\ Lett.\ B {\bf 791}, 236 (2019).

\bibitem{Peshier} A. Peshier, hep-ph/0601119 (2006).

\bibitem{LatticeMass}  O. Kaczmarek, F. Karsch, F. Zantow and P. Petreczky,
Phys. Rev. D {\bf 70} (2004) 074505; O. Kaczmarek and F. Zantow, Phys. Rev. D {\bf 71} (2005) 114510.

\bibitem{Maezawa} Yu. Maezawa {\it et al.} [WHOT-QCD Collaboration],
Phys. Rev. D {\bf 81} 091501 (2010);

\bibitem{Nakamura} A. Nakamura, T. Saito and S. Sakai, Phys. Rev. D {\bf 69},
014506 (2004).

\bibitem{Hart} A. Hart, M. Laine and O. Philipsen, Nucl. Phys. B {\bf 586},
443 (2000).

\bibitem{Bak} D. Bak, A. Karch, L. G. Yaffe, JHEP {\bf 0708}, 049 (2007).

\bibitem{ALICE_CH_v2} S.~Acharya {\it et al.} [ALICE], JHEP {\bf 1807}, 103 (2018).

\bibitem{CMS_CH_RAA}  V.~Khachatryan, {\it et al.} [CMS],  JHEP {\bf 1704}, 039 (2017).

\bibitem{CMS_CH_v2}
  A.~M.~Sirunyan, {\it et al.} [CMS], Phys.\ Lett.\ B {\bf 776}, 195 (2018).

\bibitem{ATLAS_CH_RAA} [ATLAS], ATLAS-CONF-2017-012.

\bibitem{ATLAS_CH_v2} M.~Aaboud, {\it et al.} [ATLAS],
  Eur.\ Phys.\ J.\ C {\bf 78}, 997 (2018).

\bibitem{ALICE_D_RAA} S. Acharya, {\it et al.}  [ALICE], JHEP \textbf{10}, 174 (2018).

\bibitem{ALICE_D_v2}
  S. Acharya, {\it et al.}  [ALICE], Phys.\ Rev.\ Lett. {\bf 120}, 102301 (2018).

\bibitem{CMS_D_v2}
  A. M. Sirunyan, {\it et al.}  [CMS],
  Phys. Rev. Lett. {\bf 120}, 202301 (2018).

\bibitem{CMS_JPsi} A.~M.~Sirunyan, \textit{et al.} [CMS],
  Eur.~Phys.~J.~C \textbf{78}, 509 (2018).

\bibitem{Tc}
  A.~Bazavov, {\it et al.} [HotQCD Collaboration],
  Phys.\ Rev.\ D {\bf 90}, 094503 (2014).

\bibitem{HFPuzzle}
  M.~Djordjevic, J. Phys. G \textbf{32}, S333-S342 (2006);
  M.~Djordjevic and M.~Djordjevic, Phys.\ Rev.\ C \textbf{90}, 034910 (2014).

\bibitem{Vitev:2002pf}
  I.~Vitev and M.~Gyulassy,
  Phys. Rev. Lett. \textbf{89}, 252301 (2002)

\bibitem{Renk:2010qx}
 T.~Renk, H.~Holopainen, U.~Heinz and C.~Shen,
 Phys. Rev. C \textbf{83}, 014910 (2011).

\bibitem{Betz:2014cza}
  B.~Betz and M.~Gyulassy,
  JHEP \textbf{08}, 090 (2014)
  [erratum: JHEP \textbf{10}, 043 (2014)].

\bibitem{Andres}
  C.~Andres, N.~Armesto, H.~Niemi, R.~Paatelainen and C.~A.~Salgado,
  Phys.\ Lett.\ B \textbf{803}, 135318 (2020).

\bibitem{Shi:2018lsf}
S.~Shi, J.~Liao and M.~Gyulassy,
Chin. Phys. C \textbf{42} (2018) no.10, 104104

\bibitem{JET:2013cls}
K.~M.~Burke \textit{et al.} [JET],
Phys. Rev. C \textbf{90} (2014) no.1, 014909

\bibitem{Kumar:2019uvu}
A.~Kumar, A.~Majumder and C.~Shen,
Phys. Rev. C \textbf{101} (2020) no.3, 034908.

\bibitem{LBT1} Y. He, T. Luo, X.-N. Wang, and Y. Zhu, Phys. Rev. C {\bf 91}, 054908 (2015), [Erratum: Phys. Rev. C {\bf 97}, 019902 (2018)].

\bibitem{LBT2} S. Cao, T. Luo, G.-Y. Qin, and X.-N. Wang, Phys. Rev. C {\bf 94}, 014909 (2016).

\bibitem{Xu:2014ica}
J.~Xu, A.~Buzzatti and M.~Gyulassy,
JHEP \textbf{08} (2014), 063.

\bibitem{Shi:2018izg}
S.~Shi, J.~Liao and M.~Gyulassy,
Chin. Phys. C \textbf{43} (2019) no.4, 044101.

\bibitem{Zhao:2021vmu}
W.~Zhao, W.~Ke, W.~Chen, T.~Luo and X.~N.~Wang,
[arXiv:2103.14657 [hep-ph]].

\bibitem{Werner:2012xh}
K.~Werner, I.~Karpenko, M.~Bleicher, T.~Pierog and S.~Porteboeuf-Houssais,
Phys. Rev. C \textbf{85} (2012), 064907.

\bibitem{NunesdaSilva:2020bfs}
  T.~Nunes da Silva, D.~Chinellato, M.~Hippert, W.~Serenone,
  J.~Takahashi, G.~S.~Denicol, M.~Luzum and J.~Noronha,
  Phys. Rev. C \textbf{103}, 054906 (2021)

\end{references}
\end{document}